\documentstyle[aps]{revtex} 

\draft

\title{
\begin{flushright}
{\large Yaroslavl State University\\
        Preprint YARU-HE-97/02} \\[10mm]
\end{flushright}
{\LARGE\bf Resonance neutrino bremsstrahlung $\nu \rightarrow \nu \gamma$} \\ 
{\LARGE\bf in a strong magnetic field}
\vspace*{10mm}}

\author{{\Large\bf A.A.~Gvozdev, N.V.~Mikheev } \\[2mm]
        {\large\it
             Division of Theoretical Physics, Department of Physics,} \\
        {\large\it
             Yaroslavl State University, Yaroslavl 150000, Russia} \\
        {\large\it E-mail: mikheev@yars.free.net} \\[4mm]
        {\Large\bf and} \\[4mm]
        {\Large\bf L.A.~Vassilevskaya} \\[2mm]
        {\large\it
             Moscow Lomonosov University, V-952, Moscow 117234, Russia} \\
        {\large\it E-mail: vasilevs@vitep5.itep.ru}}

\begin{document}

\maketitle

\vspace*{20mm}

\centerline{\Large\bf Abstract}

\begin{abstract}
{\large
High energy neutrino bremsstrahlung $\nu \rightarrow \nu + \gamma$ 
in a strong magnetic field ($B \gg B_s$) is studied 
in the framework of the Standard Model (SM). 
A resonance probability and a four-vector of the neutrino energy 
and momentum loss are presented. A manifestation of the neutrino 
bremsstrahlung in astrophysical cataclysm of type of a supernova 
explosion or a merger of neutron stars, such as an possible 
origin of cosmological $\gamma$-burst is discussed.}
\end{abstract}

\thispagestyle{empty}

\twocolumn[
%
%
]

Neutrino physics in media is a vigorously growing
and prospective line of investigation at the junction of 
particle physics, astrophysics and cosmology. Substance is usually 
considered as the medium. We stress that a strong magnetic field
can play role of the active medium. It will suffice to mention that
the magnetic field can influence substantially 
on neutrino electromagnetic properties (an anomalous magnetic
moment~\cite{Bor}, neutrino oscillations ~\cite{NP}, 
a magnetic catalysis of the radiative decay~\cite{GMV}). 
The investigations of this sort are, in our opinion, of principal
interest in astrophysics where gigantic neutrino fluxes and strong
magnetic fields 
($B > B_s$, $B_s = m^{2}_{e}/e \simeq 0.44 \cdot 10^{14} \, G$
is the critical, so called, Schwinger value) 
can take place simultaneously (a process of a coalescence
of neutron stars, an explosion of a supernova).
Notice that the magnetic fields at the moment of cataclysm can occur 
much more than the typical, so called, "old" magnetic fields 
of order of $B_s$ which have been observed at the surface of pulsars.
By now astrophysical knowledge of strong magnetic fields
which can be realized in the nature have changed essentially.
For example, the field strengths inside the astrophysical objects
in principle could be as high as $10^{15} - 10^{18} G$,
both for toroidal~\cite{Rud} and for
poloidal~\cite{Boc} fields. 
On the other hand, in such strong magnetic fields 
otherwise negligible processes are not only opened kinematically 
but become substantial ones as well. 

In our paper ~\cite{GMV} we have studied the massive
neutrino radiative decay $\nu \rightarrow \nu' + \gamma$ \cite{FL}
in external electromagnetic fields of various configurations,
in particular, in the strong magnetic field.
It was shown that a field-induced amplitude was not suppressed by 
smallness of neutrino masses and did not vanish even in the case
of massless neutrino as opposed to the vacuum amplitude.
The decay probability of the neutrino with energy $E_\nu < 2 m_e$
was calculated on the assumption that 
photon dispersion relation was close to the vacuum one $q^2 = 0$.

However the photon dispersion in the strong magnetic field
differs significantly from the vacuum dispersion 
with increasing photon energy~\cite{Ad,Sh1,Ts}, so 
the real photon  4-momentum can appear as a space-like  
and sufficiently large one ($\vert q^2 \vert \gg m^2_\nu$). 
In this case the phase space for neutrino transition 
$\nu \rightarrow \nu' + \gamma$ with $m < m'$ is opened also. 
It means that the decay probability of ultrarelativistic neutrino 
$\nu \rightarrow \nu' + \gamma$ becomes insensitive
to the neutrino mass spectrum due to the photon dispersion relation
in the strong magnetic field. This phenomenon results in a strong suppression
($ \sim m^2_\nu / E^2_\nu$) of the neutrino transition with flavour
violation and a diagonal process 
$\nu_l \rightarrow \nu_l + \gamma$ ($l = e, \mu, \tau$)
is realized only. Thus this diagonal radiative neutrino transition
manifests itself as neutrino bremsstrahlung, does not contain 
uncertainties associated with a possible mixing in the lepton sector 
of the SM and can lead to observable physical effects in the strong 
magnetic fields $B \gg B_s$.

In this paper we study the high energy neutrino bremsstrahlung 
in the strong constant magnetic field in the framework of the SM.
The field-induced $\nu\nu\gamma$-vertex
can be calculated using an effective four-fermion
weak interaction of the left neutrino with the electron only because
the electron is the most sensitive fermion to the external field.
By this means diagrams describing this process are reduced to an 
effective diagram with the electron in the loop (Fig.1) 


\begin{figure}[tb]

\unitlength=1.00mm
\special{em:linewidth 0.4pt}
\linethickness{0.4pt}

\begin{picture}(60.00,35.00)(-5.00,13.00)

\put(35.00,32.50){\oval(20.00,15.00)[]}
\put(35.00,32.50){\oval(16.00,11.00)[]}
\put(26.00,32.50){\circle*{3.00}}
\put(44.00,32.50){\circle*{2.00}}

\linethickness{0.8pt}

\put(11.00,42.50){\vector(3,-2){9.00}}
\put(26.00,32.50){\line(-3,2){6.00}}
\put(26.00,32.50){\vector(-3,-2){9.00}}
\put(17.00,26.50){\line(-3,-2){6.00}}

\put(36.50,39.00){\line(-3,2){4.01}}
\put(36.50,39.00){\line(-3,-2){4.01}}

\put(32.50,26.00){\line(3,2){4.01}}
\put(32.50,26.00){\line(3,-2){4.01}}

\put(23.00,28.00){\makebox(0,0)[cc]{\large $$}}
\put(47.00,28.00){\makebox(0,0)[cc]{\large $$}}
\put(18.00,42.00){\makebox(0,0)[cb]{\large $\nu(p)$}}
\put(16.00,23.00){\makebox(0,0)[ct]{\large $\nu(p')$}}
\put(34.00,45.00){\makebox(0,0)[cc]{\large $e$}}
\put(34.00,20.00){\makebox(0,0)[cc]{\large $e$}}
\put(55.00,36.00){\makebox(0,0)[cb]{\large $\gamma(q)$}}



\def\photonatomright{\begin{picture}(3,1.5)(0,0)
                                \put(0,-0.75){\tencircw \symbol{2}}
                                \put(1.5,-0.75){\tencircw \symbol{1}}
                                \put(1.5,0.75){\tencircw \symbol{3}}
                                \put(3,0.75){\tencircw \symbol{0}}
                      \end{picture}
                     }
\def\photonrighthalf{\begin{picture}(30,1.5)(0,0)
                     \multiput(0,0)(3,0){5}{\photonatomright}
                  \end{picture}
                 }


\put(44.00,32.50){\photonrighthalf}

\end{picture}

\caption{}

\end{figure}

\noindent where double lines imply that the influence of the 
external field in propagators is taken into
account exactly.

The calculation technique of this type loop diagram was 
described in detail in our paper ~\cite{GMV}.
The general expression for the $\nu\nu\gamma$-vertex has a 
cumbersome form and will be published elsewhere. We note that 
it is enhanced substantially in the vicinity of the, so called, 
photon cyclotronic frequencies similar to a vacuum 
polarization (the cyclotronic resonance in the vacuum 
polarization~\cite{Sh2}).
Here we present the amplitude of the high energy neutrino 
($E_\nu \gg m_e$) bremsstrahlung 
$\nu_l(p) \rightarrow \nu_l(p') + \gamma(q)$ 
in the case when the field strength is the greatest parameter 
$e B > E_\nu^2$. 
We stress that two eigenmodes of the photon propagation 
with polarization vectors
\begin{eqnarray}
\varepsilon _{\mu}^{(\parallel)} & = & 
\frac{ (q \varphi)_{\mu} }{ \sqrt{ 
 q^2_\perp  } } ; \; \; \; \; \;
\varepsilon _{\mu}^{(\perp)} = \frac{ (q \tilde
\varphi)_{\mu} }{ \sqrt{ q^2_\parallel } }
\label{eq:EP}
\end{eqnarray}

\noindent are realized in the magnetic field (the, so called, 
parallel ($\parallel$) and perpendicular ($\perp$) polarizations 
according to Adler's notations in the paper~\cite{Ad}).
Here $ \varphi_{\alpha \beta} = F_{\alpha \beta} / B$ and 
${\tilde \varphi}_{\alpha \beta} = \frac{1}{2} \varepsilon_{\alpha \beta
\mu \nu} \varphi_{\mu \nu} \; $ are the dimensionless tensor and dual
tensor of the external magnetic field with the strength $\vec B = (0,0,B)$;
$q^2_{\parallel}  =  ( q \tilde \varphi \tilde \varphi q ) =
q_\alpha \tilde \varphi_{\alpha\beta} \tilde \varphi_{\beta\mu} q_\mu 
= q^2_0 - q^2_3$,
$q^2_{\perp}  =  ( q \varphi \varphi q ) = q^2_1 + q^2_2$. With (\ref{eq:EP}) 
the amplitude $ M = M_{\parallel} + M_{\perp}$
is significantly simplified and can be presented in the form:
\begin{eqnarray}
M_{\parallel} & \simeq & - \frac{ e G_F }{ \sqrt{2} } \; g_v V_{\parallel} \;
\frac{ (q \varphi j) }{ \sqrt{ q^2_{\perp} } }
\label{eq:M1} \\
M_{\perp} & \simeq & - \frac{ e G_F }{ \sqrt{2} } \; \frac {1}{ \sqrt{
q^2_\parallel } } \left \lbrace g_v V_{\perp} (q \tilde \varphi j) + g_a
A (q \varphi \varphi j) \right \rbrace ,
\nonumber
\end{eqnarray}

\noindent where $M_{\parallel}$ and $M_{\perp}$ correspond to the creation 
of the $\parallel$ and $\perp$ photon modes, respectively.
Here $e > 0$ is the elementary charge; $G_F$ is the Fermi constant, 
$g_v$ and $g_a$ are the vector and axial-vector $\nu - e$ coupling 
constants in the SM, respectively;
$j_{\alpha} = 2 \bar \nu_L (p') \gamma_{\alpha} \nu_L (p)$ 
is the left neutrino current,
$V_{\parallel}$, $V_{\perp}$ and $A$ are the scalar functions of
$q^2_{\parallel}$, $q^2_{\perp}$ and $e B$.
Notice that $V_{\parallel}$ and $V_{\perp}$ 
are the known functions (see, for example, \cite{Sh2}) 
since the field-induced vacuum 
polarization is described in terms of these functions, namely:
\begin{eqnarray}
{\cal P}_{\lambda} = e^2 \pi (q^2) + e^2\,V_{\lambda}, \;\;\;
\lambda = \parallel, \perp,
\label{eq:P}
\end{eqnarray}

\noindent where $\pi (q^2)$ 
is the well-known vacuum polarization without an external field
($q^2 = q^2_{\parallel} - q^2_{\perp}$).
As for the function $A$ the result of our calculations is:

\begin{eqnarray}
A & = & \frac{ 1 }{ 4 \pi^2 } \;
\frac{ q^2_\parallel + q^2_{\perp} }{ q^2_{\perp} }
\Bigg [ i \int \limits_0^1 du \int
   \limits_0^{\infty} dt \; e^{- i \Phi } 
\nonumber \\
& \times &
\left ( m^2_e - 
\frac{ q^2_\parallel \cdot q^2 }{q^2_{\parallel} + q^2_{\perp} }
 \; \frac{ 1 - u^2 }{4} \right ) 
- e B
\Bigg ],\; 
\label{eq:A1} \\
\Phi & = & \frac{1}{e B}\left [ t \left ( m^2_e -  q^2_{\parallel}  \frac{ 1 - u^2 }{4}
\right ) +  q^2_{\perp} \; \frac{ \cos ut -
\cos  t }{ 2 \sin t }\right] , 
\nonumber
\end{eqnarray}

\noindent The important properties of functions $V_{\lambda}$ and 
$A$ are that they are singular with respect to the 
variable $q^2_{\parallel}$
in the threshold points $(q^2_{\parallel})_{thr} = {\cal E}^2_{nm}$, 
labelled by $n$, $m$:
\begin{eqnarray}
{\cal E}^2_{nm} 
= \left ( \sqrt{ m^2_e + 2 e B n } + 
 \sqrt{ m^2_e + 2 e B m } \right )^2,  
\label{eq:Enm}
\end{eqnarray}

\noindent where ${\cal E}_{nm}$ has the simple physical meaning
as transversal energy of an electron-positron pair in the magnetic field.

It is of interest for some astrophysical applications
the case of relatively high neutrino energy 
$ E_\nu \simeq 10 - 20 MeV \gg m_e$ 
and strong magnetic field $e B > E^2_\nu$ when a region of the cyclotronic 
resonance on the ground Landau level dominates in this process.
The functions $V_{\perp}$~\cite{Sh2} and $A$ are easily calculated in this 
region:

\begin{eqnarray}
A \simeq - V_{\perp} \simeq  \frac{e B \; m_e \, 
e^{{- q^2_{\perp}}/{2 e B}} }
{2 \pi \sqrt{4 m^2_e - q^2_{\parallel}}  }.
\label{eq:A2}
\end{eqnarray}

\noindent As it is seen from ~(\ref{eq:A2}) the amplitude  
$ M_{\perp}$~(\ref{eq:M1}) 
contains the enhancement due to the square-root singularity which is 
connected with the cyclotronic resonance on the ground Landau level
($q^2_\parallel \rightarrow 4 m_e^2$). As for $V_{\parallel}$,
it has not such an enhancement and the dispersion relation for
$\parallel$ photon mode is close to the vacuum one. So, the
amplitude $M_{\parallel} \sim (q \varphi j) \sim q^2$ 
is small due to collinear kinematics of the process  
with creation of ${\parallel}$ photon mode.
The square-root singularities on higher Landau levels are 
not realized under the condition $e B > E^2_\nu$. We note that not only 
function $A$  has singular behaviour in the vicinity
of the threshold point $q^2_\parallel = 4 m_e^2$, but the vacuum 
polarization ${\cal P}_{\perp}$~(\ref{eq:P}) as well.
This means that many-loops quantum corrections are of
great importance in the vicinity of the resonance. As analys shows
taking account of these radiative corrections reduces to the 
renormalization of the $\perp$ photon mode wave function: 
\begin{eqnarray}
\varepsilon^{(\perp)}_{\mu} 
\rightarrow \sqrt{Z} \;\varepsilon^{(\perp)}_{\mu}, \,\,\,\,
Z = \left ( 1 - 
\frac{\partial {\cal P}_{\perp}}
{\partial q^2_\|}\right ) ^{-1},
\label{eq:EQ}
\end{eqnarray}

\noindent and leads to an additional factor $ \sqrt{Z}$ in 
the amplitude $M_{\perp}$~(\ref{eq:M1}).

The probability of the neutrino bremstrahlung $\nu \rightarrow \nu \gamma$ 
can be obtained by integration over the
phase space with taking account of the photon $q^2 - {\cal P}_{\perp} = 0$ 
in the vicinity of the resonance:
\begin{eqnarray}
4 m^2_e - q^2_{\parallel} & \simeq & \alpha^2 
\left ( \frac{2 e B}{q^2_{\perp}} \right )^2 m^2_e 
\ll q^2_{\perp}.
\nonumber
\end{eqnarray}

\noindent This leads to the dependence of photon energy on the 
photon momentum $ q_0 \simeq  \sqrt{ q^2_3 + 4 m^2_e }$
(the photon created can move only along
the magnetic field with the velocity 
$v = dq_0 / dq_3 \simeq q_3 / \sqrt{q^2_3 + 4 m^2_e}$ ).
The main contribution to the probability is determined 
from the vicinity of the resonance point  
$q^2_\parallel = 4 m_e^2$ and can be presented in the form: 

\begin{eqnarray} 
W & \simeq & \frac{\alpha G^2_F}{8 \pi^2 }(g^2_v + g^2_a) 
(e B)^2 E_\nu \sin^2{\theta}\, F(\lambda), 
\label{eq:W1} \\
F(\lambda) & = & 1 - \frac{\lambda}{2} + \frac{\lambda^2}{3}
- \frac{5 \lambda^3}{24} + \frac{7 \lambda^4}{60} + O(\lambda^5),
\nonumber \\
\lambda & = & \frac{E_\nu^2 \sin^2{\theta}}{e B} < 1, 
\nonumber 
\end{eqnarray}

\noindent where $\theta$ is the angle between the vectors 
of the magnetic field strength ${\vec B}$ and the momentum of
the initial neutrino ${\vec p}$. 

We stress the principal importance of taking account of the photon
wave function renormalization~(\ref{eq:EQ}). If one ignored this 
factor the decay probability would be strongly overstated:
\begin{eqnarray}
W \simeq \left( \frac{1}{\alpha} \right)
\left (\frac{5 G^2_F m_e^5}{256 \pi^2} \right)(g^2_v + g^2_a)
\left(\frac{E_\nu}{m_e}\right)^7 \sin^8{\theta}.
\label{eq:WS}
\end{eqnarray}

It is of more practical importance for some astrophysical 
applications to calculate the four-vector of the neutrino energy 
and momentum loss in the process of the neutrino bremsstrahlung:

\begin{equation}
Q_\mu \, = \, E_\nu \int d W \cdot q_\mu.
\end{equation}

\noindent The physical meaning of this vector is that 
$Q_0/E_\nu$ is the mean energy loss in a unit time and
$\vec Q/E_\nu$ is the momentum loss in a unit time.
We have obtained the expression for the vector $Q_\mu$: 
\begin{eqnarray}
Q_\mu & = & \frac{1}{4}\;E_\nu W \Big ( F_1(\lambda) 
(p \tilde \Lambda)_\mu - 2 F_2(\lambda) (p \Lambda)_\mu 
\nonumber \\
& - & \frac{2 g_v g_a}{g^2_v + g^2_a} F_1(\lambda)  
(p \tilde\varphi )_\mu \Big ),
\label{eq:Q1} \\
F_1(\lambda) & = & 1 - \frac{2\lambda}{3} + \frac{5\lambda^2}{12}
- \frac{7 \lambda^3}{30} + \frac{7 \lambda^4}{60} + O(\lambda^5),
\nonumber \\
F_2(\lambda) & = & 1 - \lambda + \frac{5\lambda^2}{6}
- \frac{7 \lambda^3}{12} + \frac{7 \lambda^4}{20} + O(\lambda^5).
\nonumber 
\end{eqnarray}

To illustrate the applications of the results we have obtained
we estimate below possible effects of the neutrino 
bremsstrahlung in astrophysical cataclysm of type of a 
supernova explosion or a merger of neutron stars under the condition
that a compact remnant has for some reasons 
extremely strong magnetic field, 
$B \simeq 10^{16} - 10^{18} G$. 
We assume that neutrinos of all species with the typical mean 
energy $\bar E_\nu \simeq 20 MeV$ 
are radiated from a surface of a neutrinosphere in such a process. 
Using (\ref{eq:Q1})
we estimate the neutrino energy loss:

\begin{eqnarray}
{\cal E} & \simeq & 10^{50}\;  
\left (\frac{{\cal E}_{tot}}{ 10^{53} erg} \right )
\left (\frac{B}{10^{17}G} \right )^2
\label{eq:E} \\
& \times &
\left (\frac{\bar E_\nu}{20 MeV}\right )
\left (\frac{ R}{10 km}\right ) erg,
\nonumber
\end{eqnarray}

\noindent where ${\cal E}_{tot} \simeq 10^{53}$ erg
is the total neutrino radiation energy ~\cite{JR};
$B$ is the magnetic field strength in the vicinity of
the neutrinosphere of radius $R$. 
Neutrino bremsstrahlung $\gamma$-quanta are captured 
by a strong magnetic field and propagate along the 
field~\cite{Sh2}. Thus the mechanism of
significant power "pumping" of polar caps of a magnetized 
remnant can take place.
This phenomenon can result in reemission within a narrow cone 
$\Omega/4 \pi \ll 1$ along the magnetic moment of the remnant.
So, if an external absorbing envelope is absent in such an
extraordinary astrophysical cataclysm (very strong magnetic fields,
gigantic neutrino fluxes) the reemission process 
will be an observable effect. Namely, such a reemission in the 
magnetic field could appear as an anisotropic $\gamma$-burst with 
the duration of order of the neutrino emission time and of the 
typical energy ${\cal E}_{4\pi}^{burst} \sim 10^{51}$ erg 
in terms of $4 \pi$-geometry.

\vspace{5mm}

The authors thank V.A.~Rubakov and V.M.~Li\-pu\-nov for fruitful 
discussions, M.E.~Sha\-posh\-ni\-kov and A.V.~Kuz\-ne\-tsov for
useful critical remarks.
The work of N.V.~Mikheev  was supported by a Grant N~d104
from the International Soros Science Education Program.
The work of L.A.~Vassilevskaya  was supported by a fellowship
of INTAS Grant 93-2492-ext and was carried out within the research
program of International Center for Fundamental Physics in Moscow.


\begin{references}
\bibitem{FL}
We stress that $\nu$ and $\nu'$ are the neutrinos
eigenstates with definite masses, but not the neutrinos with
definite flavours.
\end{references}
\end{document}